\documentstyle[aps, floats, epsf, twocolumn]{revtex}

\begin{document}

\title{Antiferromagnetism from phase disordering of a d-wave superconductor}

\author{Igor F. Herbut}

\address{Department of Physics, Simon Fraser University, 
Burnaby, British Columbia, \\
 Canada V5A 1S6\\} \maketitle

\begin{abstract}
The unbinding of vortex defects in the superconducting condensate with  
d-wave symmetry at $T=0$ is shown to lead to the insulator with
incommensurate spin-density-wave order. The transition
is similar to the spontaneous generation of the chiral mass in the
three dimensional quantum electrodynamics. Possible relation to recent
experiments on underdoped cuprates is discussed.
\end{abstract}

\vspace{10pt}

Common feature of all high-temperature superconductors is that undoped they
are Mott insulators with antiferromagnetic order \cite{millis}.
The central theme of the theories 
of cuprate superconductivity has therefore been to establish  the connection
 between the insulating and the superconducting phase.
Most of the work followed the usual route that suggests starting from
the non-superconducting, in this case, Mott insulating phase,
and trying to understand how it becomes superconducting.
This approach was spectacularly successful for the conventional (low-$T_c$)
 superconductors, in part because
the non-superconducting phase was a well understood metallic Fermi liquid.
In cuprates, however, one does not enjoy this luxury, and
the Mott insulator is strongly interacting and notoriously resistant
to simple theoretical understanding. This suggests one should look
for alternative points of view that may be better adapted to the
problem at hand. Since experimentally the superconducting phase seems to be 
a rather standard BCS-like d-wave state, one strategy would be to take
this as a vantage point for further exploration of the cuprates
phase diagram \cite{fisher}, \cite{franz}. Particularly 
interesting is the underdoped region, where experiments show a large pseudogap
for spin excitations, and the superconductor-insulator transition
at low temperatures.

 In this Letter I subscribe to the dual approach advocated
 above and show that the d-wave superconducting state 
(dSC) at $T=0$ has an instability towards the insulator with the
incommensurate spin density wave (SDW) order. Using the 
Franz-Te\v sanovi\' c transformation  
I derive the low-energy theory for the coupled
system of d-wave quasiparticles and fluctuating vortices. Upon integration
over vortices the theory takes the form of the (anisotropic) 2+1 
dimensional quantum electrodynamics (QED3) for  two Dirac
four-component spinors, which are related to the nodal quasiparticles
by a singular gauge transformation,  and are minimally
coupled to the transverse gauge field \cite{franz}.
First, I show that the role of the coupling constant (or  the 
"charge")  in this gauge theory is at $T\neq 0$
played by the thermodynamic
fugacity of the vortex system. In the superconducting
phase the "charge" is therefore zero,
vortices are bound into pairs,
and the gauge field is decoupled from the fermions.
In the non-superconducting phase, on the other hand,
the fugacity is finite, and the gauge field now mediates
a long-range interaction between the Dirac fermions. The main result is 
that, at $T=0$ where the role of fugacity is played
by the condensate of vortex loops, this interaction leads
to an instability towards the incommensurate SDW 
order, through a condensed-matter equivalent of the chiral symmetry breaking
phenomenon \cite{appelquist}. The $T=0$ transition from the 
dSC into the SDW may be understood therefore as an instability
of the gapless nodal fermionic excitation in presence of free
topological defects towards the formation of
bound states. Possible connections between
recent neutron scattering, ARPES, and STM experiments are discussed
in light of this result.

What follows rests on two postulates: 1)
that there is a d-wave superconducting
state in the phase diagram 
with sharp gapless quasiparticle excitations,
and 2) that the amplitude of the superconducting
order parameter may be assumed
finite and inert much below the high pseudogap temperature $T^*$,
so that the only other relevant excitations are the
topological defects in its phase (vortices and antivortices
at $T\neq 0$, or vortex loops at $T=0$). The first postulate
is supported by the microwave \cite{bonn}, and the ARPES
experiments \cite{arpes}, and the second by the measurements
of the frequency dependent conductivity \cite{corson}
and the Nernst effect in the pseudogap regime \cite{xu}.
I begin by constructing the continuum, low-energy theory for the nodal
quasiparticles in the d-wave state, using a 
different representation than in \cite{fisher}, \cite{franz}. The
quasiparticle Hamiltonian is 
\begin{eqnarray}
H_{qp} = T \sum _{\vec{k}, \sigma, \omega_n }[  (i\omega_n -\xi_{\vec{k}
}) c^{\dagger}
_{\sigma}(\vec{k},\omega_n ) c_{\sigma} (\vec{k},\omega_n ) \\ \nonumber 
-\Delta(\vec{k}) c^{\dagger}
_{\sigma}(\vec{k},\omega_n ) c^{\dagger}_{-\sigma} (-\vec{k},- \omega_n )
+ c. c. ],
\end{eqnarray}                                                       
where $\Delta(\vec{k})$ has the usual d-wave symmetry, and 
two spatial dimensions (2D) are assumed. $c$ and $c^{\dagger}$ are the
electron operators, $\sigma=\pm$ labels spin, and $\omega_n $ are
the fermionic Matsubara frequencies. In my units $h=c=e=1$. Next,
introduce {\it two four-component} Dirac spinors
\begin{eqnarray}
\Psi^{\dagger}_{1(2)} (\vec{q},\omega_n ) = (c^{\dagger}_+ (\vec{k},\omega_n ),
c_- (-\vec{k}, -\omega_n ), \\ \nonumber
c^{\dagger}_+ (\vec{k}-\vec{Q}_{1(2)}, \omega_n ),
c_- (-\vec{k}+\vec{Q}_{1(2)}, -\omega_n ) ), 
\end{eqnarray}
where $\vec{Q}_{1(2)}= 2\vec{K}_{1(2)}$ is the wavevector that connects
the nodes within the diagonal pair 1(2).
For the spinor 1, $\vec{k}=\vec{K}_1+\vec{q}$,
with $|\vec{q}| \ll |\vec{K_1}| $ (see Fig. 1), and 
analogously for the second pair.
One has $\xi_{\vec{k}} = \xi_{- \vec{k}}$,  and
{\it near} the nodes,  
$\xi_{\vec{k}}= - \xi_{ \vec{k} - \vec{Q}_{1(2)} }$, and 
$\Delta_{\vec{k}}= - \Delta_{ \vec{k} - \vec{Q}_{1(2)} }$,
for $\vec{k}\approx \vec{K}_{1(2)}$.  Retaining only the low-energy
modes in (1), and linearizing the spectrum as $\xi_{\vec{k}} = v_f q_x$ 
and $\Delta_{\vec{k}} = v_\Delta q_y $, one arrives at the continuum
field theory
\begin{eqnarray}
S[\Psi] = \int d^2 \vec{r} \int_0 ^ {\beta} d\tau \bar{\Psi}_1 [ \gamma_0 \partial_\tau
+ \gamma_1 v_f \partial_x + \gamma_2 v_\Delta
\partial_y ] \Psi_1  +  \\ \nonumber
( 1 \rightarrow 2, x\rightarrow y, y \rightarrow x), 
\end{eqnarray}
 where $\bar{\Psi} = \Psi \gamma_0$, and the matrices
 $\gamma_0 = \sigma_1 \otimes I$,
 $\gamma_1 = \sigma_2 \otimes \sigma_3$, and $\gamma_2 = - \sigma_2 \otimes
 \sigma_1$ satisfy the Clifford algebra
 $\{ \gamma_\mu, \gamma_\nu \} = 2\delta_{\mu,\nu}$.
 Here $\vec{\sigma}$ are the Pauli matrices, and the 
 coordinate system has been rotated as in Fig. 1.

   Next, assume that the transition out of the dSC in the underdoped
regime is due to unbinding of the topological defects.
This raises a rather non-trivial
question of how to properly couple the vortex degrees of freedom to
quasiparticles \cite{fisher}. Fortunately, this has recently been
elegantly solved by Franz and Te\v sanovi\' c
 \cite{franz}, \cite{senthil}. Their idea is to split the phase of the
order parameter $\Delta(\vec{r}, \tau) = |\Delta|
\exp{i ( \phi_s (\vec{r},\tau)+ \phi_r (\vec{r},\tau)) }$,
where $\phi_{s(r)}$
is the singular (regular) part of the phase in presence of vortices, 
into two contributions, $\phi_s +\phi_r  = \phi_A + \phi_B $, with 
\begin{eqnarray}
\nabla \times \nabla \phi_{A}(\vec{r},\tau) = 2\pi
\sum_{iA =1}^{N_{A}}
 q_{iA} \delta(\vec{r}-\vec{r}_{iA}(\tau) ), 
\end{eqnarray}
where $q_{iA} =\pm 1$ is the unit vorticity of a 
($hc/2e$) vortex (antivortex)
defect, tracked by its coordinate in the  imaginary time 
$\vec{r}_i (\tau)$, and analogously for vortices in B.
Here $\nabla=(\partial_{\tau}, \partial_x,\partial_y )$.
The division of vortices into two groups A and B is
at this point arbitrary, and $\phi_r$ is to be equally split between $\phi_A$
and $\phi_B$.  By making the singular gauge transformation
in (1) from electrons into electrically {\it neutral} fermions
$ c_{+(-)} \rightarrow c_{+(-)} \exp {i\phi_{A(B)} } $, 
one immediately discovers that there is a hidden gauge field in the problem, 
$a_\mu = \frac{1}{2} \partial_\mu (\phi_A - \phi_B)$, $\mu=0,1,2$,
that enters the theory (3) via {\it minimal} coupling
$\partial_\mu \rightarrow \partial_\mu - i a_\mu$. The 
regular part of the phase $\phi_r $ cancels in $\vec{a}$,
which is entirely due to vortices.  $\phi_r$ is contained
in the second, Doppler gauge field, 
$v_\mu = \frac{1}{2} \partial_\mu (\phi_A + \phi_B)$, which 
enters the theory for the neutral fermions precisely  as the 
true electromagnetic gauge field would.
Gauge invariance protects $\vec{a}$ from becoming massive from
the integration over fermions, while such a protectorate does not exist
for $\vec{v}$. Power counting implies then that the coupling of fermions
to $\vec{v}$ is irrelevant, and may and
will therefore be dropped hereafter.

\begin{figure}
\centerline{\epsfxsize=5cm \epsfbox{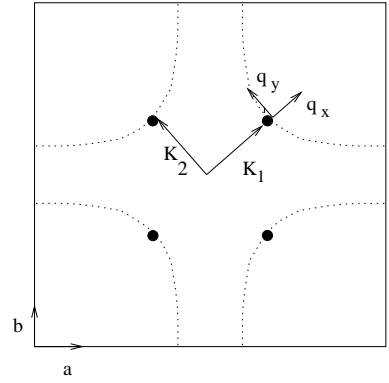}}
\vspace*{1em}

\noindent
\caption{The wavevectors $\vec{K}_1$, $\vec{K}_2$, and $\vec{q}$.}
\label{f1}
\end{figure}

It was argued in \cite{franz} that although $\vec{a}$ can not become
massive from fermions, it should be massive if vortices are bound.
Next I present a simple but a rigorous derivation of the dynamics
of the gauge-field $\vec{a}$ at $T\neq 0$, where one can get away with 
the neglect of the quantum fluctuations, which supports
this insight. Assume a collection
of $N_+ (N_-)$ vortices (antivortices) at the positions
$\{ \vec{r}_i \}$. The energy of the (classical) vortex system is
\begin{equation}
H_v = \frac{1}{2} \sum_{i=1}^N  q_i q_j v(\vec{r}_i - \vec{r}_j), 
\end{equation}
where $v(\vec{r})\approx   -\ln |\vec{r}|$, at large distances,
and $N=N^+ + N^- $.  The partition function of the coupled
system of quasiparticles and vortices can be written as
\begin{equation}
Z= \int D[\Psi] e^{-S[\Psi, \vec{a} ]}  Z_v, 
\end{equation}
where $Z_v$ is the grand-canonical classical 
partition function of the vortex system (2D Coulomb plasma)
\begin{equation}
Z_v = \sum_{N_A^- , N _A^+ , N_B^- , N_B^+}
\frac{y^N
\int \prod_{i=1} ^{N} d\vec{r}_i
e^{-\frac{H_v}{T}} }
{2^N  N_A^+ ! N_B^+ ! N_A^- ! N_B ^- ! } , 
\end{equation}
where $N^{+(-)} = N_A^{+(-)} + N_B^{+(-)}$, and $y$ is the bare
vortex fugacity. To preserve the $\sigma\rightarrow - \sigma$
symmetry in the original Hamiltonian in $Z_v$ I average
over {\it all} possible divisions of vortices and antivortices
into groups A and B. This ensures that on average
there is an equal number
of vortices (and antivortices) in both groups. Next, introduce the
vorticity densities in $Z_v$ by inserting the unity 
\begin{equation}
1= \int D[\rho_{A}] \delta( 
\rho_{A} (\vec{r}) - \sum_{i=1}^{N_{A}} q_{iA}\delta ( \vec{r}
- \vec{r}_{iA})),   
\end{equation}
 and similarly for B. The gauge field then  becomes
\begin{equation}
(\nabla\times \vec{a}(\vec{r}) )_{\tau}  = 2\pi (\rho_A (\vec{r}) 
- \rho_B(\vec{r}) ), 
\end{equation}
in the transverse gauge $\nabla\cdot\vec{a}=0$. Subindex $\tau$ denotes
the $\tau$-component.
$\vec{v}$ is defined the same way except with the plus sign between
$\rho_A$ and $\rho_B$. Performing then the Gaussian integrations over
$\rho_A$, $\rho_B$, and $\vec{v}$, and the summations in the Eq. (7)
{\it exactly}, yields $Z_v = \int D[ \Phi_+, \Phi_- , \vec{a}]
\exp(-S_v[ \Phi_+, \Phi_- , \vec{a}])$,  with 
\begin{eqnarray}
S_v [ \Phi_+, \Phi_- , \vec{a}]   =
\int d^2 \vec{r} [2T (\nabla \Phi_+(\vec{r}))^2 + \\ \nonumber
\frac{i}{2\pi}
\Phi_- (\vec{r}) (\nabla\times \vec{a}(\vec{r}) )_{\tau} 
 - 2 y \cos(\Phi_+(\vec{r}) ) \cos(\Phi_-(\vec{r}) )].
\end{eqnarray}
Real fields $\Phi_{\pm}= \Phi_A \pm \Phi_B $ are the Lagrange multipliers
introduced to enforce the constraints in the Eq. (8) \cite{negele}.

 The partition function of the coupled system of 
d-wave quasiparticles and vortices at $T\neq 0$ is therefore
$Z=\int D[ \Psi, \Phi_+, \Phi_-, \vec{a}]\exp ( - S[\Psi, \vec{a}] -
S_v [ \Phi_+, \Phi_- , \vec{a}]) $ with
\begin{equation}
S[\Psi, \vec{a}] = \sum_{i=1}^{F}
\int d^2 \vec{r} \int_0 ^{\beta}
d\tau \bar{\Psi}_i  \gamma_\mu (\partial_\mu - i a_\mu)
\Psi_i , 
\end{equation}
with $F=2$,
and the $x\leftarrow\rightarrow y$ exchange  of the coordinates for the
$i=2$ component is assumed.
I have also set $v_f = v_{\Delta}=1$ here for simplicity.
The Dirac field $\Psi$ represents the neutral (gauge-transformed) fermions, and
$S_v$ is given by the Eq. (10). This is my first result. It has several
remarkable features. First, if one turns off the coupling to fermions 
(by taking, formally, the quenched limit $F=0$),
the integration over $\vec{a}$ in (10) simply 
enforces $\Phi_- \equiv 0$. $S_v$ reduces then to the standard
sine-Gordon theory, which is known to provide the 
correct description of the Kosterlitz-Thouless transition
 \cite{minhagen}. More importantly for our purposes,
for $F=0$ one also finds
\begin{equation}
\langle (\nabla\times\vec{a}(\vec{r}))_{\tau}
(\nabla\times\vec{a}(\vec{r}'))_{\tau} 
\rangle = \langle y  \rangle
\delta(\vec{r}-\vec{r}'),  
\end{equation}
where $\langle y \rangle  = y (2 \pi) ^2
 \langle \exp ( i \Phi_+) \rangle$, with the
average to be taken over $S_v$ with $\Phi_- \equiv 0$.  
$\langle y \rangle $ may be recognized as 
 the {\it thermodynamic}, or the renormalized,  fugacity
of the vortex system \cite{minhagen}.
The integration over the fields $\Phi_+$ and $\Phi_-$ 
is thus equivalent to reducing the action (10) to 
\begin{equation}
S_v\rightarrow \int d^2 \vec{r} \frac{ (\nabla\times\vec{a})
_{\tau} ^2 }{ 2 \langle  y \rangle } 
\end{equation}
in the partition function $Z$.
In the dielectric phase of the vortex system the field 
$\Phi_+$ is massless, and consequently $\langle y \rangle  =0$
\cite{minhagen}, so the gauge-field asymptotically 
decouples from the fermions. Quasi-particles become sharp excitations
in the dSC, in agreement with the ARPES \cite{arpes}
and the microwave measurements \cite{bonn}. In the
non-superconducting phase, on the other hand, vortices are free,
$\Phi_+$ becomes massive,  and $\langle y \rangle \neq 0 $.
This has profound consequences for the fermions, 
as I discuss shortly. 

  At $T=0$ quantum fluctuations need to be included, as the topological
defects in $2+1$ dimensions become vortex loops \cite{sudbo}.
It seems clear on physical grounds, however,
 that after the integration over the
loops, (apart from the inherent anisotropy,) 
 the form of the action for the gauge field should remain similar to 
the Eq. (13), except that $(\nabla\times \vec{a}) _{\tau}^ 2/(2\langle y
\rangle ) \rightarrow   (\nabla\times \vec{a})^ 2 / ( 2 \langle y \rangle)$.
This can also be derived on a lattice,
where one finds that the role of the 
coupling $\langle y \rangle$ at $T=0$ is assumed by the {\it dual }
order parameter that becomes finite only when there are infinitely large
loops in the system, and which is tantamount to the loss of phase coherence
\cite{herbut}.
This way one finally arrives at the QED3 with the full Maxwell term
for the transverse gauge-field $\vec{a}$ as the relevant low-energy theory.

QED3 has been extensively studied by field theorists
as a non-trivial toy model exhibiting the phenomena of dynamical
symmetry breaking and confinement
\cite{appelquist}. In particular, it has been established that
for the number of Dirac fields $F<F_c$ the interaction with the gauge-field
is strong enough to spontaneously generate the so-called chiral 
mass for fermions.
I will demonstrate that the chiral mass in the theory (11)
is nothing but the SDW order parameter. First, to acquire some sense for  
the chiral instability consider the fermion propagator.
Neglecting the vertex and the wave-function renormalizations,  it
can be written as $G^{-1}(p) = i\gamma_\nu p_\nu + \Sigma (p)$,
where the self-energy satisfies the self-consistent equation
\begin{equation}
\Sigma(q) =\langle y \rangle  \gamma_\mu \int \frac{d^3 \vec{p}}{(2\pi)^3 }
\frac{ D_{\mu \nu} ( \vec{p}-\vec{q} ) \Sigma (p) }
{p^2 + \Sigma^2 (p) } \gamma_\nu,
\end{equation}
with $\vec{q}= (\omega, q_x, q_y)$.
The gauge-field propagator in the transverse gauge is
$D_{\mu \nu} = (\delta_{\mu \nu} - \hat{p}_\mu \hat{p}_\nu )/(p^2 + \Pi (p) )$,
where $\Pi (p)$ is the self-consistently determined polarization.
Assuming $m=\Sigma (0)\neq 0 $ gives  \cite{appelquist} 
\begin{equation}
\Pi (q) = \frac{\langle y \rangle F }{2\pi} (m
+\frac{q^2 - 4 m ^2 }{2q} \sin^{-1} \frac{q}{
\sqrt{q^2 + 4 m ^2 } } ). 
\end{equation}
When this is
inserted into the Eq. (14), it can be shown that
there is a solution with a finite $m$ only when
$F<F_c$, with $F_c = 32/\pi^2$ \cite{appelquist}. More  elaborate 
calculations that fully include the wave-function renormalization and
the vertex corrections confirm this result, and yield
$F_c \approx 3 $ \cite{maris}. Simulations
on the lattice version of the QED3 \cite{kocic} also
find $3<F_c<4$, in agreement with the analytical estimates.

By reversing the transformations that led to the QED3 
the reader can convince himself that the mass term
for neutral fermions is equivalent to the low-energy part  of
the following addition to the electronic Hamiltonian (1): 
\begin{equation}
m T \sum_{\vec{k},\sigma , 
\omega_n , \vec{q}=\pm \vec{Q}_{1,2}} \sigma 
 c^{\dagger}_\sigma (\vec{k} + \vec{q}, \omega_n )
 c_\sigma (\vec{k}, \omega_n),
\end{equation}
so that the chiral mass may be identified
with the SDW order parameter (or the staggered potential)
along the spin z-axis, and  at the wavevectors $\vec{Q}_{1,2}$.  This is,
of course, why the particular construction of the Dirac field
was made in the first place. Note that: 1) the SDW order is induced already
at an infinitesimal vortex fugacity, but it is rather {\it 
weak}, $m\approx \langle y \rangle  /
\exp (2\pi/ \sqrt{ (F_c/F) -1}) $ for $F\approx F_c$ \cite{appelquist},
and 2) neutral fermions are bound (confined) at large distances  
in the SDW, by the weak logarithmic potential (provided by the fact
that $\Pi(q)= \langle y \rangle F q^2 /(6 \pi m)$ for $q \ll m $). 
With some anisotropy ($v_f \neq v_\Delta$), the global symmetry of the
massless theory is only $U(2) \times U(2)$, so the mass term reduces each
$U(2)$ to $U(1)\times U(1)$. The two broken
generators per Dirac field rotate the "cos-SDW" in the Eq. (16) into
either the similar "sin-SDW", or into the phase-incoherent state with
"$d+ip$" pairing between the neutral fermions. SDW is therefore the only
state that can be obtained by the unbinding of vortex defects and that respects
parity. In the isotropic limit the massless theory recovers the
full $U(4)$ symmetry and additional broken symmetry states become
available, like, remarkably, the stripe-like charge density waves
parallel to $a$ or $b$ axis \cite{herbut}.

Once it is realized that unbinding of vortices leads to the SDW order,
it becomes natural to wonder what the nature of vortices inside
 the dSC could be. From the perspective of this work
it seems more than plausible that vortex cores are actually in the 
insulating phase, so that by approaching  half-filling one lowers
the core energy. In this picture the chemical potential  should be
related to the bare vortex fugacity, which when too large
leads to the proliferation of defects, in analogy to the
Berezinskii-Kosterlitz-Thouless transition \cite{minhagen}.
The idea of SDW in vortex cores finds some experimental
support in the recent STM \cite{renner} and the neutron scattering studies
\cite{aeppli}, as well as in the mean-field calculations \cite{ting}.
Furthermore, the present work suggests that the superconductor-insulator 
transition should be accompanied by the appearance of the 
incommensurate SDW correlations at the wave vectors $\vec{Q}_{1,2}$, 
with the incommensurability {\it increasing} with doping. This  
is consistent with the recent neutron scattering experiments
\cite{fujita}
on the underdoped LaSrCuO close to the superconducting transition.
Finally, the d-wave pseudogap should continuously evolve into the
insulating state, except for the gap that should develop at the nodes.
This also seems in agreement with the observations \cite{ronning}.

To summarize, I showed how liberating topological defects in the
d-wave superconductor at $T=0$ leads to 
the incommensurate SDW, which is then expected to
continuously evolve into the commensurate antiferromagnet
close to half-filling. Near the transition the SDW order
is inherently weak due to the relative closeness of the two
flavor QED3 to its chiral critical point at $F_c \approx 3$.
The SDW transition temperature near the superconductor-insulator
transition may therefore be expected to be much lower than the
corresponding superconducting
$T_c$ on the other side of the transition, not in contradiction
with the known topology of the cuprates phase diagram.
The issue of a quantum-disordered (deconfined) ground state
in this approach reduces to whether $F_c$ may in fact be smaller than two.
It has recently been argued that $ F_c \leq 3/2$ \cite{appel}, in spite
of the results of virtually every calculation violating this bound. Recent
numerical study on larger systems than in \cite{kocic},
for example,  find a small, but definitely a
finite mass at $F=2$ \cite{alexandre}. More interesting possibility is that
$F_c$ may depend on some additional parameter in the
theory.  For example, one may speculate that a 
large anisotropy $v_f / v_\Delta\sim 10 $, which is certainly present
in cuprates, could affect $F_c$. This way
$F_c$ would become doping dependent, which could open a route for 
the $T=0$ deconfined phase in between the dSC and the SDW.
Further examination of this possibility, together with the issue of the
order of transition, effects of disorder and finite temperature, and the
relations to other theoretical approaches to the high-$T_c$ problem
will be discussed in the future longer publication \cite{herbut}.

The author is grateful to S. Dodge, M. Fisher, S. Sondhi,
S. Sachdev, Z. Te\v sanovi\' c, and W.-C. Wu for useful discussions
and criticism, and particularly to D. J. Lee for pointing out the
exactness of the Eq. (10). This work was supported by the NSERC of Canada
and the Research Corporation.

\end{document}